\documentclass[12pt]{article}
\usepackage{psfig}
\usepackage{amsfonts}
\begin{document}

\def\lesssim{\mathrel{\mathpalette\vereq<}}
\def\gtrsim{\mathrel{\mathpalette\vereq>}}
\makeatletter
\def\vereq#1#2{\lower3pt\vbox{\baselineskip1.5pt \lineskip1.5pt
\ialign{$\m@th#1\hfill##\hfil$\crcr#2\crcr\sim\crcr}}}
\makeatother

\newcommand{\rem}[1]{{\bf #1}}
\newcommand{\gev}{\rm GeV}
\newcommand{\mev}{\rm MeV}
\newcommand{\kev}{\rm keV}
\newcommand{\ev}{\rm eV}
\newcommand{\cm}{\rm cm}
\renewcommand{\thefootnote}{\fnsymbol{footnote}}
\setcounter{footnote}{0}
\begin{titlepage}
\begin{center}
                                                                                  
%
                                                                                   
%
   \hfill    
UCB-PTH-00/43\\                                                      %
   \hfill    LBNL-47234\\     
   \hfill    
UW/PT-00-17\\                                               %
   \hfill    
hep-ph/0101138\\                                                     %
   \hfill    \today 
\\                                                   

   \vskip 
.5in                                                                    
{\Large \bf Inelastic Dark Matter\footnote{This work was supported in part by the 
U.S. Department of Energy under Contracts DE-AC03-76SF00098, in part 
by the National Science Foundation under grant PHY-95-14797.}                         %
}                                                                              
   \vskip 
.50in                                                                   
   David Smith$^{1}$ and Neal Weiner$^{2}$

   \vskip 
0.05in                                                                  
   $^{1}${\em Department of 
Physics\\                                                   %
        University of California, Berkeley, California 
94720}                     %

   \vskip 
0.05in                                                                  

and                                                                            
   \vskip 
0.05in                                                                  
   {\em Theoretical Physics 
Group\\                                               %
        Ernest Orlando Lawrence Berkeley National 
Laboratory\\                    %
        University of California, Berkeley, California 
94720}                     %
       \vskip 0.05in
           
           $^{2}${\em Department of Physics\\
           University of Washington, Seattle, Washington 98195, USA}

   \vskip 
.5in                                                                    

\end{center}                                                                   
   \vskip 
.5in                                                                    

\begin{abstract}                                                               
Many observations suggest that much of the matter of the universe is 
non-baryonic. Recently, the DAMA NaI dark matter direct detection 
experiment reported 
an annual modulation in their event rate consistent with a WIMP relic. 
However, the Cryogenic Dark Matter Search (CDMS) Ge experiment excludes most 
of the region preferred by DAMA. We demonstrate that if the dark 
matter can only scatter by making a transition to 
a slightly heavier state ($\Delta m \sim 100\kev$), 
the experiments are no longer 
in conflict. Moreover, differences in the energy spectrum of nuclear
recoil events could distinguish 
such a scenario from the standard WIMP scenario. Finally, we 
discuss the sneutrino as a 
candidate for inelastic dark matter in supersymmetric theories.
   
\end{abstract}                                                                 

\end{titlepage}                                                                

\renewcommand{\thepage}{\arabic{page}}                                         

\setcounter{page}{1}                                                           

\renewcommand{\thefootnote}{\arabic{footnote}}                                 

\setcounter{footnote}{0}                                                       

\setcounter{footnote}{0}
\setcounter{table}{0}       
\setcounter{figure}{0}
\section{Introduction}

A central task of modern cosmology is to determine what 
the universe made of. A number of observations suggest that the 
bulk of the matter in the universe is not luminous \cite{Trimble:1987ds}. 
Direct searches for baryonic matter in the form of massive compact 
halo objects (MACHOs) cannot account for the matter that seems 
necessary to explain these observations \cite{Alcock:2000ph}.

An alternative explanation is that weakly interacting massive 
particles (WIMPs) exist copiously in the halo of our 
galaxy but only rarely interact with ordinary
matter \cite{Lee:1977ua}. Candidate WIMPs from particle theory
include the axion and 
the lightest supersymmetric particle (LSP) in supersymmetric theories 
with R-parity conservation.

Numerous experiments have been set up in attempts to directly 
detect WIMPs \cite{Baudis:2001ph,Abusaidi:2000wg,Bernabei:2000qi}. 
The two which are sensitive 
to the smallest spin independent cross sections are the CDMS Ge 
experiment \cite{Abusaidi:2000wg} and the DAMA NaI experiment \cite{Bernabei:2000qi}. 
Recently, DAMA reported the presence of a signal consistent with a 
WIMP at better than $4 \sigma$. When interpreted as a standard WIMP 
with spin independent interactions, CDMS rules out nearly all of the 
DAMA $3\sigma$ preferred region at $90 \% $
confidence and all of it at $84 \% $
confidence. Attempts to reconcile these experiments using spin 
dependent interactions have been shown to be in gross conflict with 
indirect detection experiments and previous direct searches \cite{Ullio:2000bv}. 

In this paper, we will show that a simple modification to the 
properties of the dark matter particle can change the kinematics of 
the scattering sufficiently to reconcile the two experiments.
In particular, we explore the possibility of inelastic dark
matter:  relic particles that cannot scatter elastically off of nuclei.
The outline of the paper is as follows: we begin by comparing the 
details of the two experiments and give a naive argument why inelastic 
dark matter can reconcile them. In section two we  
explicitly calculate the event rate at CDMS and DAMA taking into 
account the inelasticity of the scattering. In section three we  
use this calculation to study what differences can arise relative
to the elastic case and to examine whether 
there are regions of parameter space
that give a signal at DAMA but a null result at CDMS. 
In section four we discuss how inelastic dark matter 
could arise from a massive complex scalar split into two approximately 
degenerate real scalars, or from a Dirac fermion split into two 
approximately degenerate Majorana fermions. We also present a specific
model, featuring a real component of the sneutrino as the dark matter,
in which the mass splitting required to reconcile DAMA and CDMS arises
naturally. In section five we discuss 
direct detection possibilities at future experiments. 

\subsection{CDMS and DAMA}
If we are to understand the DAMA signal as evidence of dark matter, but 
simultaneously accept the null result of CDMS, we must reconsider 
some basic element of the WIMP hypothesis.
Before we address such a modification, we should understand the 
differences between the DAMA and CDMS experiments.

The DAMA experiment utilizes a set of NaI crystals at the Gran Sasso 
National Laboratory of INFN to search for WIMPs. The basic premise of 
the experiment is that if WIMPs are present in the galaxy, as the 
galaxy rotates we feel a ``wind'' of WIMPs which will
scatter elastically off of the target nuclei. As the Earth moves 
in its orbit about the sun, the flux and velocity distribution (as seen 
by a terrestrial observer) vary. Rather than attempt to directly 
discriminate signal events against background, the DAMA experiment 
seeks to measure this modulation. There are two basic controls to 
this experiment.  First, the signal phase must coincide with the 
Earth's motion in the solar system, which moves maximally with the 
galactic rotation on June 2, and maximally against on December 2. The 
second requirement is that the signal must lie dominantly in the lowest energy 
bins - a characteristic signal of WIMP scattering.

In contrast, CDMS uses a smaller Ge target, but has excellent 
background 
rejection capable of distinguishing nuclear recoils from electron 
scatterings for scattering energies greater than $10 \kev$. As a 
consequence their limits are comparable to those that would have been 
expected from a null DAMA result.

Exclusion plots are typically given in the $m_{\chi}-\sigma_{n}$ 
plane, where $m_{\chi}$ is the mass of the candidate and $\sigma_{n}$ 
is the scattering cross section per nucleon. Implicit is the 
assumption that there are no great modifications in the scattering 
process between the two experiments.

However, if the dark matter cannot scatter elastically, then kinematical 
effects substantially distinguish the experiments.  Consider
two states, $\chi_{-}$ and 
$\chi_{+}$, with $\chi_{+}$ only slightly heavier than $\chi_{-}$, 
such that $\chi_{-}$ can only scatter by transitioning to $\chi_{+}$.
It is a simple kinematical constraint that $\chi_-$ 
can only scatter inelastically off of a nucleus with mass $m_N$ if
\begin{equation}
        \delta < \frac{\beta^{2} m_{\chi} m_{N}}{2(m_{\chi}+m_{N})},
        \label{eq:constraint}
\end{equation}
where $\delta$ is the mass splitting between $\chi_-$ and $\chi_+$.
The possibility of evading direct detection by having a large enough
splitting $\delta$ was pointed out in \cite{Hall:1998ah}.  Here we
focus on the fact the constraint of equation (\ref{eq:constraint})
becomes increasingly severe as $m_N$ is decreased. 
Since iodine has an atomic number of 127, while 
germanium has an atomic number of 73, we have the prospect of a 
situation where particles will scatter at DAMA but not at CDMS.
For $\beta c \approx 220{\rm km/s}$ (a typical dark matter 
particle velocity), and $m_{\chi} = 100 \gev$, the limits are $11 \kev$ for 
CDMS and $15 \kev$ for DAMA. If the mass splitting $\delta$ were $13 
\kev$, such a particle would be visible to DAMA but not CDMS.

Of course, in the halo of the galaxy there is a distribution 
of velocities, so the calculation is not as simple as 
we have just illustrated. 
In the full calculation, we will find that the values of $\delta$
relevant for reconciling the experiments are somewhat larger than
15 keV, and that the window for $\delta$ has a size $\sim 50 - 100$ keV
rather than $\sim 5$ keV. 

\section{Direct Detection Rates}
In this section we review the standard calculation of event rates at
direct detection experiments \cite{Freese:1988wu}.  
The differential rate per unit detector
mass is given by 
\begin{equation}
{dR \over dE_R}= N_T {\rho_\chi \over m_\chi} \int_{v_{min}} \!dv \; v
f(v)
{d\sigma \over dE_R}.
\label{eq:rate1}
\end{equation}
Here $E_R$ is the recoil energy of the target nucleus, $N_T$ is the
number of target nuclei per unit mass, $\rho_\chi$ is the local
density of dark matter particles of mass $m_\chi$, ${d\sigma \over
  dE_R}$ is the differential cross section for relic-nucleus
scattering, and $v$ and $f(v)$ are the relic speed and speed
distribution function in the detector rest frame.  We take
$\rho_\chi=.3$ GeV/cm$^3$.

Because we are interested in spin-independent scattering, the
differential cross section may be written  
\begin{equation}
{d\sigma \over dE_R}={m_N \over 2 v^2}{\sigma_n\over
  \mu_n^2}{\left(f_p Z+f_n(A-Z) \right)^2\over f_n^2}
F^2(E_R),
\label{eq:cs1}
\end{equation}
where $m_N$ is the nucleus mass, $\mu_{n}$ is the reduced mass of
the relic-nucleon system, $f_n$ and $f_p$ are the relative coupling
strengths to neutrons and protons, and $\sigma_{n}$ is the
relic-neutron cross section at zero momentum transfer, in the
elastic $(\delta=0)$ limit.  We use the Helm form factor \cite{form}
\begin{equation}
F^2(E_r)=\left({3 j_1(q r_0)\over q r_0}\right)^2 e^{-s^2q^2},
\end{equation}
with $q=\sqrt{2 m_N E_R}$, $s=1$ fm, $r_0=\sqrt{r^2-5s^2}$, and
$r=1.2A^3$.

We assume a standard Maxwell-Boltzman distribution for the relic
velocities in the galactic rest frame, with a root-mean-squared
velocity $v_{rms}=\sqrt{3\over 2}v_0$, where we take $v_0=220$ km/s
to be the rotational speed of the local standard of rest (LSR).
In our calculation we take the escape velocity to infinity for
simplicity, when one really should take 
$v_{esc}\simeq 650$ km/s.  By doing so we
overestimate the signal for large values of the mass splitting
$\delta$.  For a 100 GeV relic, this is a 10\% effect at CDMS for
$\delta=100$ keV and a factor of two effect for $\delta=150$ keV.
Because iodine is heavier than germanium, the effect is far milder at
DAMA, roughly 10\% at $\delta=150$ keV. 

The earth's speed relative
to the galactic rest frame is
\begin{equation}
v_e={v_\odot} + v_{orb} \cos\gamma \cos\left(\omega(t-t_0)\right).
\end{equation}
Here $v_\odot=v_0+12$ km/s, $v_{orb} =30$ km/s, $\omega =2\pi$/year,
$t_0 \simeq$ June 2nd, and $\cos \gamma=.51$.  Defining the
dimensionless variables $\eta=v_e/v_0$ and $x_{min}=v_{min}/v_0$, 
performing the velocity integration in (\ref{eq:rate1}),
and applying (\ref{eq:cs1}), one obtains
\begin{equation}
{dR \over dE_R}= {N_T m_N \rho_\chi \over 4 v_0 m_\chi}F^2(E_R)
{\sigma_n\over
  \mu_n^2}{\left(f_p Z+f_n(A-Z) \right)^2\over f_n^2}
\left({{\rm
  erf}(x_{min}+\eta)-{\rm erf}(x_{min}-\eta)\over \eta}\right).
\label{eq:rate2}
\end{equation} 
For the DAMA detector, one should take into account there being two
species of target nuclei with different quenching factors.

Often one considers the case where $f_n=f_p$
(so that the rate is proportional to $A^2$), and presents results 
in the $m_\chi$ - $\sigma_n$ plane.  Below we will be particularly
interested in models in which the scattering is dominated by vector
interactions arising from $Z$ boson exchange, giving $f_n/
f_p=-(1-4\sin^2\theta_W)\simeq -.08$ (and yielding a rate that is instead nearly proportional to
$(A-Z)^2$).
In all of our calculations we take
this value for $f_n/f_p$.

The differential rate
of equation (\ref{eq:rate2}) depends on the mass splitting parameter
$\delta$ through $x_{min}$, which is given by 
\begin{equation}
x_{min}={1\over v_0}\sqrt{1\over 2 m_N E_R}\left({m_N E_R \over
\mu}+\delta\right),
\end{equation}  
where $\mu$ is the reduced mass of the relic-nucleus system.  A
non-zero $\delta$ increases the minimum relic speed required to produce
a given nuclear recoil energy.  In the following section we explore
potential consequences for direct
detection signals arising due to this modification.

\section{Signals at CDMS and DAMA}
\label{sec:sigs}
\begin{figure}
  \centerline{
    \psfig{file=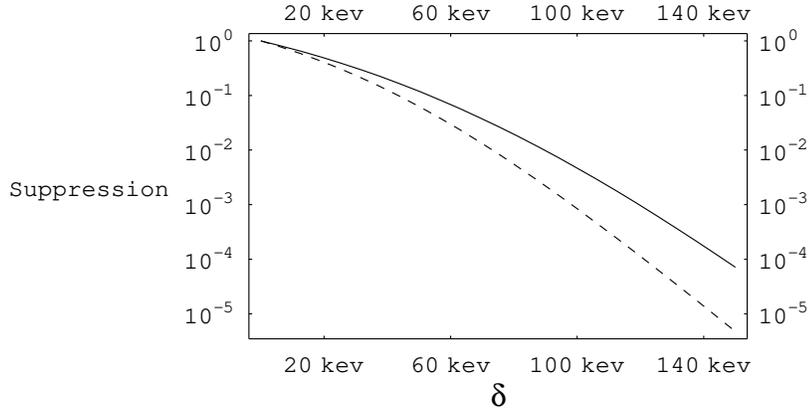,width=0.8\textwidth,angle=0} }
        \caption{Ratio of total events in iWIMP scenario to ordinary 
        WIMP as a function of splitting $\delta$ for DAMA (solid 
        line) and CDMS (dashed), with $m_{\chi}=50\gev$. 
                For DAMA we have integrated the total events in the 
        $2-10 \kev$ energy region, while for CDMS we have integrated 
        in the $10 - 100 \kev$ region. For large $\delta$ ($>$ 100
        keV), 
the finite 
        value of the the galactic escape velocity can become 
        important, yielding larger suppressions than shown.  This
        effect is stronger for CDMS than for DAMA.}
        \label{fig:relsupp}
\end{figure}

Before we study whether there are regions of parameter space that 
are consistent with both DAMA and CDMS, it is worthwhile to 
investigate the differences arising when compared with the 
elastic case. We have seen that for a given velocity of 
dark matter particle, it might be that only DAMA is be able to
detect the particle, and not CDMS. Given the 
distribution of relic velocities, we can now determine what effect the
inelasticity has on the full 
signal integrated over all velocities. The simplest 
quantity to consider is the level at which the 
signal\footnote{For our purposes here, we will consider the signal to 
be the events falling in the $10 \kev - 100 \kev$ bins for CDMS and 
$2 \kev - 10 \kev$ for DAMA.} is suppressed when compared with the 
elastic case. We plot these suppressions for CDMS and DAMA 
in figure \ref{fig:relsupp}.

We can easily see that our basic intuition is borne out. The greater 
the splitting between $\chi_{-}$ and $\chi_{+}$, the greater the 
suppression for CDMS compared to that of DAMA. Since the CDMS 
excluded 
region only just covers the DAMA preferred region, even a factor of a few can 
dramatically improve the consistency of the experiments. 

\begin{figure}
  \centerline{
    \psfig{file=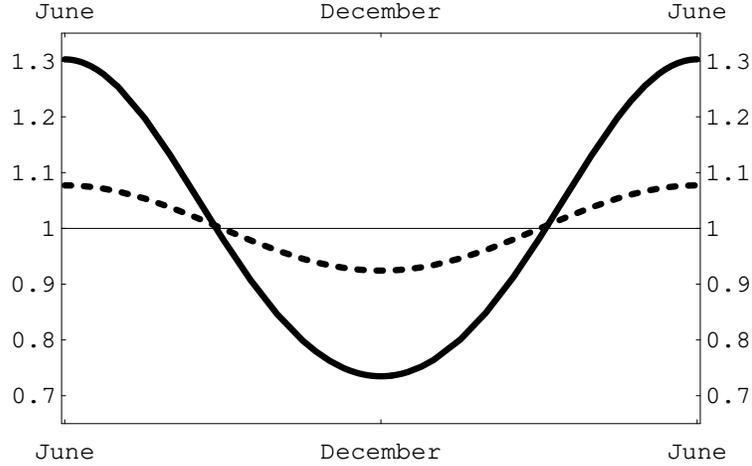,width=0.65\textwidth,angle=0} }
        \caption{Annual modulation of event rate with average 
        normalized to one in the inelastic 
WIMP scenario (solid line) and standard 
        WIMP scenario (dashed), with $\delta=100 \kev$ and 
        $m_{\chi}=50\gev$.}
        \label{fig:relmod}
\end{figure}

However, the relative suppression is not the only relevant quantity 
because DAMA is not sensitive to the total flux, but rather to 
the modulation of the flux. Because of the inelasticity,  
DAMA only sees those particles on 
the high tail of the Maxwellian distribution. Consequently,
a small modulation in 
the average velocity can lead to much higher modulation for a given 
signal when compared with the elastic case. This effect is 
demonstrated in figure \ref{fig:relmod}. The combination of these two 
effects results in DAMA having significant regions of sensitivity 
that are inaccessible to the existing Ge experiments.

The DAMA signal can be decomposed into background, unmodulated signal 
and modulated signal as
\begin{equation}
        \mu_{k} = b_{k}+ S_{0,k}+ S_{m,k} \cos (\omega t),
        \label{eq:DAMAsignal}
\end{equation}
where $k$ indexes the energy bin of each piece of the total measured 
events 
$\mu$. Recently, the DAMA collaboration published its best fit values 
of $S_{0,k}$ and $S_{m,k}$ for the energy bins 
$2-3\kev$, $3-4\kev$, $4-5\kev$ and $5-6\kev$ \cite{Bernabei:2000qi}. 
It is tempting to fit 
the inelastic scattering case to these values, but to do so would be 
misleading. These best fit values are derived assuming the energy 
spectrum and relative size of the modulated piece to be given by 
the known relations for an elastically scattering WIMP. We have 
already seen in figure \ref{fig:relmod} that the standard WIMP and
inelastic WIMP cases can lead to very different predictions for the
relative size of the modulated piece, 
so any fit to the published best-fit values would not be 
rigorous.

\begin{figure}
  \centerline{
    \psfig{file=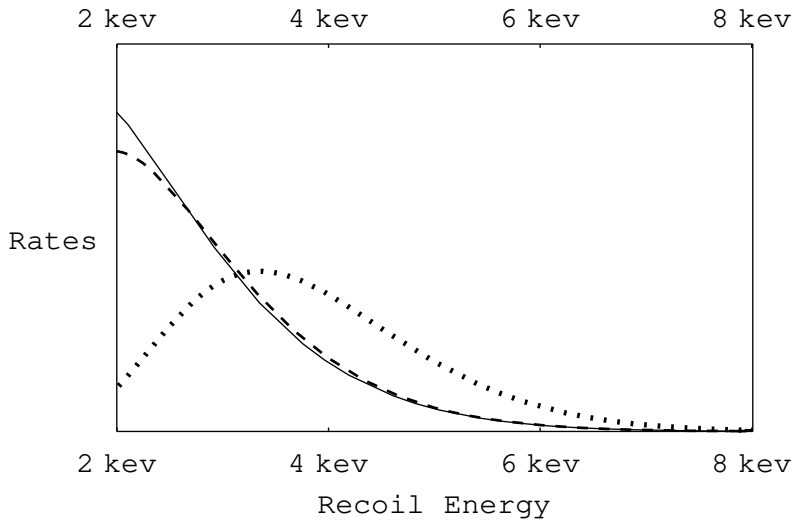,width=0.65\textwidth,angle=0} }
        \caption{Normalized modulation ($S_{m}$) as a function of 
energy for 
        ordinary WIMP scenario (solid), 
inelastic WIMP scenario with $\delta = 100 
\kev$ (dashed), and inelastic WIMP scenario with $\delta= 150 \kev$ 
        (dotted), all with $m_{\chi}=60 \gev$.}
        \label{fig:modspectrum}
\end{figure}

Another, potentially more significant reason that we cannot use the standard 
WIMP $S_{m,k}$ values comes from changes in the energy spectrum of 
the events. Because the scattering is inelastic, the total number of 
events may not rise exponentially at low energy. In other cases, the 
spectra will be nearly identical. As examples we 
compare in figure \ref{fig:modspectrum} the expected WIMP spectrum of
the modulation signal to the spectrum in the inelastic 
WIMP scenario for two values of $\delta$. The potential 
differences revealed in figure \ref{fig:modspectrum}
make it possible to fit only to the model independent data recently 
published. We will discuss the details of this fit shortly. 

\begin{figure}
  \centerline{
    \psfig{file=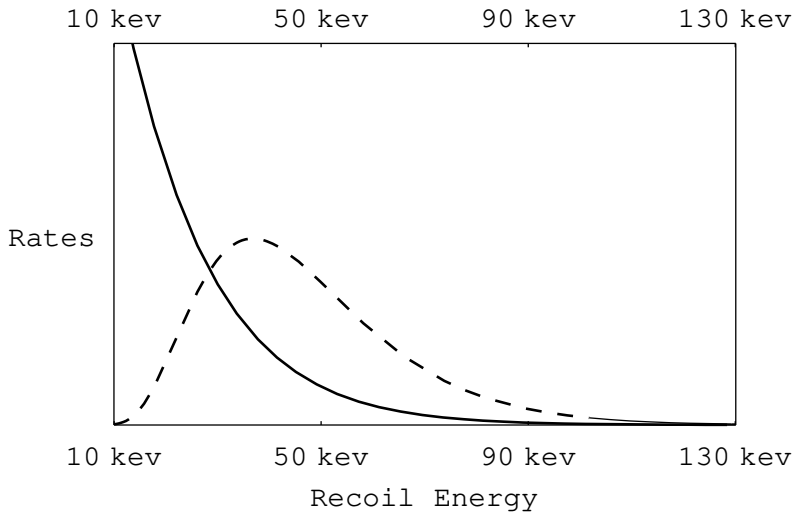,width=0.65\textwidth,angle=0} }
        \caption{Normalized spectrum of events at CDMS for ordinary WIMP 
(solid) and 
        inelastic WIMP (dashed) with $\delta = 100 \kev$, both with 
$m_{\chi}=50 \gev$.}
        \label{fig:cdmsspec}
\end{figure}

These spectrum differences carry over to germanium experiments. As we 
show in figure \ref{fig:cdmsspec}, the changes can again be 
significant, and can again alter the interpretation of the 
experimental data. For instance, in the elastic case 
one expects an exponential rise 
in the number of relic scattering events for lower energies.
Were CDMS to see many 
events in the $40-60 \kev$ bin, but essentially an absence of events 
below $40 \kev$, this would be inconsistent with an elastic dark matter 
signal, but not with an inelastic dark matter signal. 
Again, we do not perform a rigorous fit to the CDMS data as this 
would require an
ability to accurately simulate the correlation of multiple scatterers 
with single scatterers, which we lack. 

For the purposes of generating allowed regions, we will thus use the 
following limits: for DAMA, we will use the published model 
independent modulation in the $2-6 \kev$ bins of $0.088 \pm 0.02
{\rm counts/day/kg}$ \cite{Bernabei:2000qi} and consider the three sigma region to be allowed. 
DAMA 
claims not to have modulation in the higher energy bins. Although the 
measured modulation for energies above $6 \kev$ is not published, we 
will take an upper limit of $0.003 {\rm counts/day/kg}$, which we consider 
quite reasonable given the errors on the best fit values for the 
higher energy bins. 
For CDMS, we will require a predicted mean of fewer than six total 
events, consistent with the published limits \cite{Abusaidi:2000wg}.

DAMA has also reported null results arising from a pulse shape 
analysis (PSA) of a portion of their NaI data \cite{Bernabei:1996vj} 
and of data from an experiment with Xe $(A=129)$ \cite{Bernabei:1998ad}. 
Using the pulse shape, they can 
discriminate signal from background, and place a limit on the total 
number of events. Both of these studies affect the elastic WIMP 
preferred region for DAMA. Extracting rate limits from tables and plots of 
\cite{Bernabei:1996vj} and 
 \cite{Bernabei:1998ad}, we find the Xe studies have
the dominant impact on our allowed regions.
For the Xe experiment we require the signal to 
be less than $0.7$ ${\rm counts/day/kg}$ for the $13-15\kev$ bin, 
$0.25$ ${\rm counts/day/kg}$ for the $15-20\kev$ bin, $0.15$ ${\rm 
counts/day/kg}$ for the $20-25\kev$ bin, and $0.075$ 
${\rm counts/day/kg}$ for the 
$25-30\kev$ bin, consistent with published limits \cite{Bernabei:1998ad}.

\begin{figure}
  \centerline{
    \psfig{file=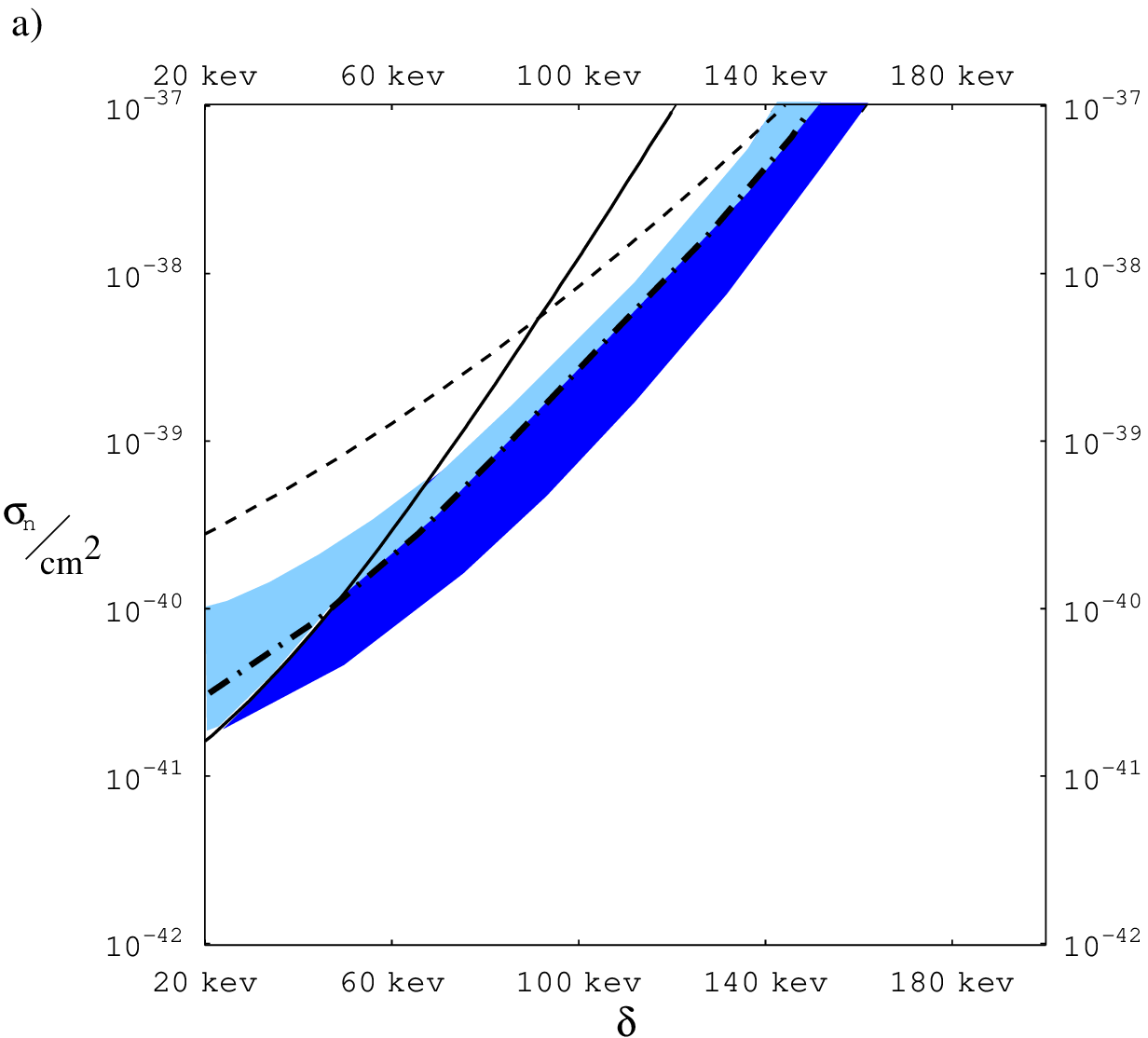,width=0.5\textwidth,angle=0}
    \psfig{file=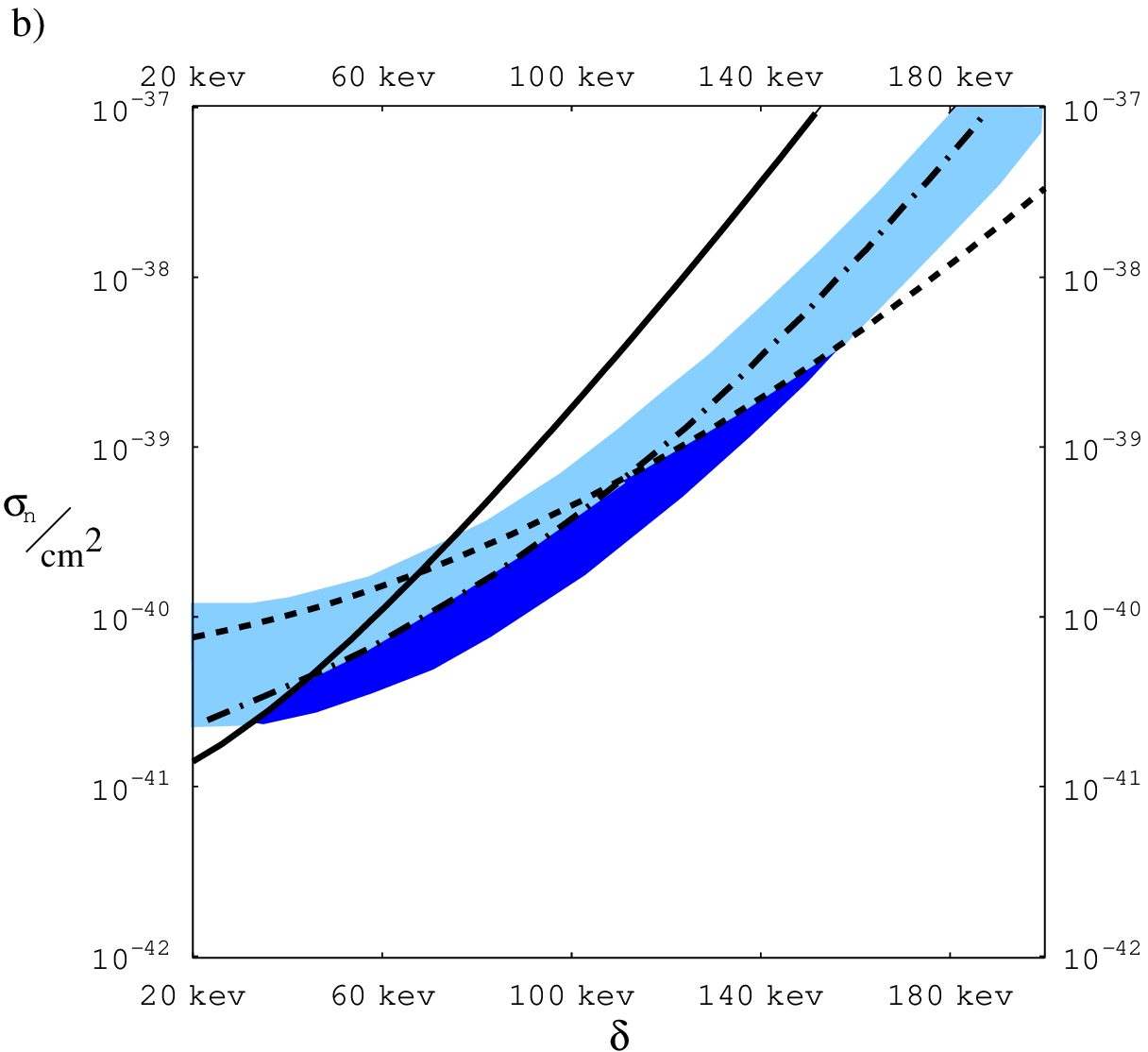,width=0.5\textwidth,angle=0} }
\centerline{
        \psfig{file=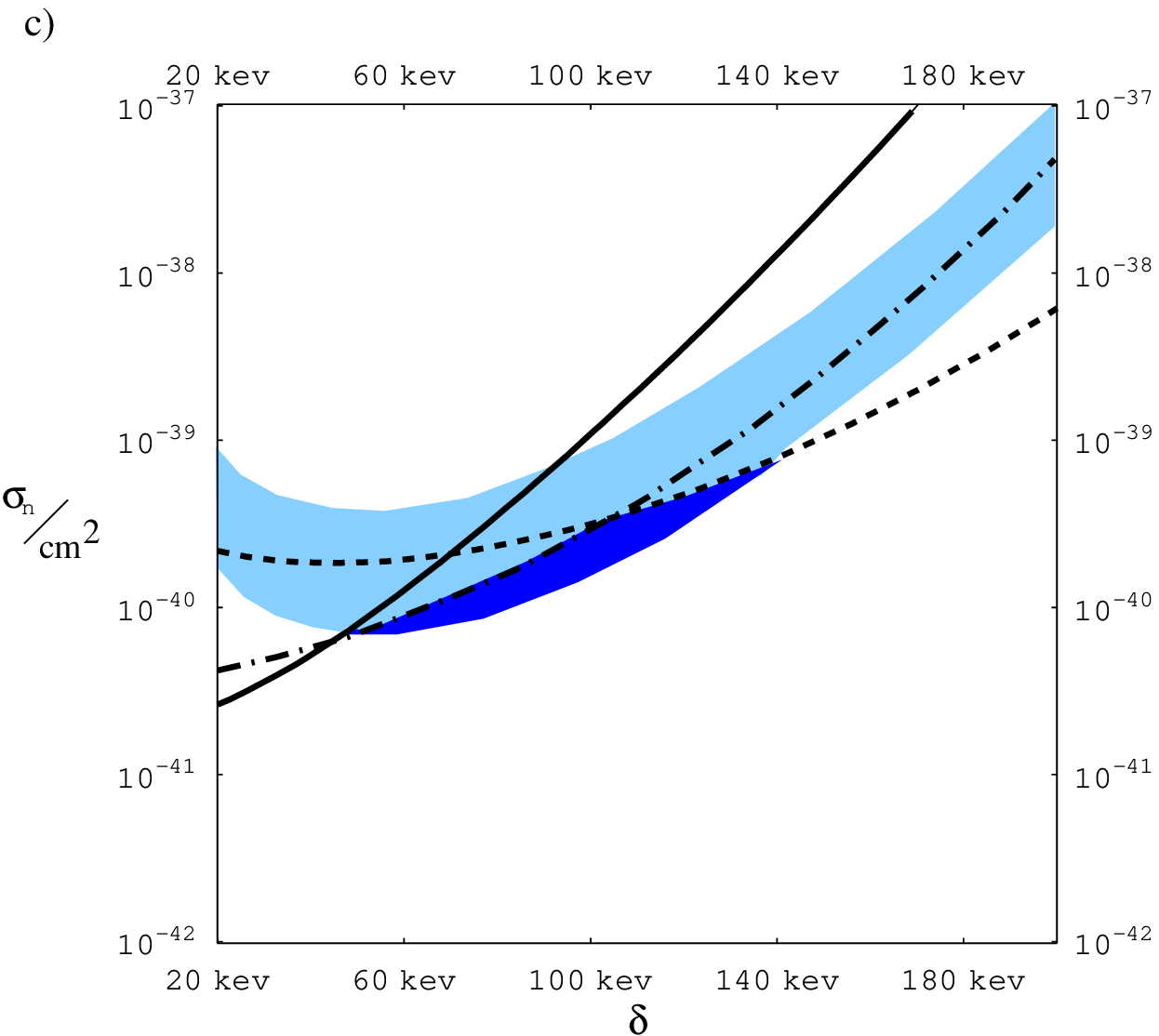,width=0.5\textwidth,angle=0} }
                \caption{Regions satisfying both DAMA and CDMS constraints in 
       the $\delta-\sigma_{n}$ plane, for (a) $m_{\chi}=50\gev$, (b) 
        $m_{\chi}=100\gev$, (c) $m_{\chi}=300\gev$. In each plot, the 
        shaded region has an integrated signal in the $2-6\kev$ energy 
        range 
        consistent with the DAMA $3\sigma$ region. 
        The solid line gives the CDMS constraint and the dashed line 
        gives the limit from an assumption of 
                the absence of signal in the high energy 
        bins at DAMA. The dot-dashed line gives the upper bound arising 
        from Xe pulse shape analysis limits. 
                The dark shaded region satisfies all constraints 
        simultaneously.}
        \label{fig:regions}
\end{figure}

We show the allowed regions subject to these constraints for various 
values of $m_{\chi}$ in figure \ref{fig:regions}. As expected, there
are broad regions that fit the DAMA data and which are not excluded by CDMS.
It is important to note that our qualitative results are not very 
sensitive to the details of the criteria used to determine 
what signals are consistent with experiment. The general 
features of figure \ref{fig:regions} remain essentially intact even 
if we are more conservative in our estimates of the allowed counts at 
CDMS, or of the accuracy of the measured modulation at DAMA. 

As an explicit example, let us consider the point 
$m_{\chi}= 70\gev$, $\delta= 105 \kev$ and $\sigma_{n} = 5 \times 
10^{-40}\cm^{2}$. Here 
the modulation is quite consistent with the DAMA best fit point, but 
CDMS has only an expected signal of $0.5$ events, and the Xe pulse 
shape analysis constraints are evaded. A comparison 
between the inelastic point and the DAMA best fit values for the 
elastic case is given in table \ref{tb:splits}.
\begin{table}
 \begin{center}
  \begin{tabular}{||l|c|c||} 
          \hline
Energy  & iWIMP & DAMA  \cr
\hline
$2-3 \kev$ & $0.021$ & $0.023 \pm 0.006$ \cr
$3-4 \kev$ & $0.014$ & $0.013 \pm 0.002$ \cr
$4-5 \kev$ & $0.007$ & $0.007 \pm 0.001$ \cr
$5-6 \kev$ & $0.003$ & $0.003 \pm 0.001$ \cr
\hline
\end{tabular}
\label{tb:specpoint}
\end{center}
\caption{Binned signal rates for an inelastic WIMP with $m_{\chi}= 70\gev$, 
$\delta= 105 \kev$ and $\sigma_{n} = 5 \times 
10^{-40}\cm^{2}$, compared with the DAMA best fit values for a 
standard WIMP. CDMS would have seen an expected 0.5 events.}
\label{tb:splits}
\end{table}

\subsection{Cosmological Uncertainties}
\label{subsection:cosmun}

Unlike the ordinary WIMP scenario, the only inelastic WIMPs that scatter in 
existing experiments are those on the high end of the Maxwell-Boltzman 
velocity distribution. As such, there is greater uncertainty in the 
precise values of $\sigma_{n}$ that fit the data than for an 
ordinary WIMP.

In particular, there is significant uncertainty in the dispersion 
velocity $v_{rms}$ and in the local halo velocity $v_{0}$. Although 
these uncertainties are $O(10\% )$, the effects can be amplified 
because of the presence of the exponential in the distribution. We 
have investigated these effects 
and found that the preferred cross sections can 
can shift by as much as a factor of three for $m_{\chi}=100\gev$ and 
a factor of seven for $m_{\chi}=50\gev$. Likewise the local 
density $\rho_\chi$ is uncertain to a factor of approximately two, and
moreover, the presence of substructure in the halo of the galaxy can 
lead to amplifications of the local density relative to the average halo 
density by a factor of three or more \cite{Moore:1999wf}. 

Finally, we must restate that we have assumed a Maxwell-Boltzmann 
distribution, which arises in the isothermal sphere model of dark 
matter. Changes to the velocity profile of the dark matter can have 
significant effects on the modulation for standard WIMPs 
\cite{Vergados:2000rh}, and inelastic dark matter potentially is even 
more sensitive to these changes. Such uncertainties are 
difficult to quanitify and we do not discuss them further.

Altogether these uncertainties can amount to a change in the preferred
$\sigma_{n}$ values, but we should emphasize that the sizes of the
regions of parameter space that yield consistency between DAMA and
CDMS do not change dramatically.

\section{Models of Inelastic Dark Matter}
Up to this point, we have considered inelastic dark matter as an 
interesting phenomenological possibility, but have not addressed 
how such particles might arise in a reasonable model. 
One possibility is that the relic particle is a real scalar, so that
its vector coupling to nuclei is forbidden by Bose symmetry.
Consider a complex scalar 
%
$\phi ={1\over \sqrt{2}}(a+ib)$ coupled to an Abelian gauge 
field $A_{\mu}$.  Its vector interaction comes from
\begin{equation}
|D_{\mu}\phi|^2\supset -g A_{\mu} (a \partial^{\mu} b- b \partial^{\mu} a).
\end{equation}
That is, the real scalars $a$ and $b$ couple to each other, but
neither couples to itself.  

These real scalars are degenerate if the only mass term is
$-m^2 |\phi|^2$, but introducing a small additional mass term
%
$-{\Delta^{2}} \phi^2 + h.c.$, 
splits this degeneracy\footnote{Of 
course, $\Delta$  violates gauge invariance, 
and can only arise once the gauge symmetry of 
the theory has been broken.}. 
If $m$ is roughly $100$ GeV, and we want
a splitting $\sim 100$ keV, then we require 
$\Delta^{2} \sim (100\; \mev)^{2}$.
In the model of section \ref{subsection:model}, which features a real
component of a sneutrino as the dark matter, this scale for $\Delta$
arises naturally.

Before discussing this model, we note that
the inelastic dark matter could instead be
fermionic.  Consider a Dirac fermion 
$\psi=\left(\begin{array}{cc}\eta  & \overline{\xi}\end{array}
\right)$
that has vector and axial-vector couplings to quarks:
\begin{equation}
\overline{\psi}\gamma_{\mu}(g_V'+g_A' \gamma_5)\psi \;\overline{q}
\gamma^{\mu}(g_V+g_A \gamma_5) q.
\label{eq:fermioncoupling}
\end{equation}
Assuming for simplicity that the various $g$'s are of comparable size,
the largest contribution to the low-energy scattering of $\psi$ off of
nuclei will come from the vector-vector piece, which will yield an
amplitude that scales roughly as the number of nucleons.
The axial-axial piece yields a smaller spin-dependent contribution 
that lacks this enhancement, while the vector-axial pieces vanish in
the extreme non-relativistic limit.  

Now suppose that in addition to a Dirac mass $\sim 100$ GeV for
$\psi$, the Lagrangian also contains a very small Majorana mass term
${\delta \over 2}(\eta \eta+\overline{\eta}\; \overline{\eta})$, with
$\delta\sim 100$ keV.  Then the Majorana fermion mass eigenstates are
\begin{eqnarray}
\hspace{2in}\chi_1 \simeq {i \over \sqrt{2}}(\eta-\xi)\hspace{1.in} m_1=m-\delta \\
\hspace{2in}\chi_2 \simeq {1 \over \sqrt{2}}(\eta+\xi)\hspace{1.in} m_2=m+\delta.
\end{eqnarray}
The vector current essentially couples $\chi_1$ to $\chi_2$, with only 
a  small additional piece $\sim \delta /m$ coupling each mass
eigenstate to itself:
\begin{equation}
\overline{\psi} \gamma_\mu \psi \simeq i(\overline{\chi}_1
\overline{\sigma}_{\mu} \chi_2-\overline{\chi}_2
\overline{\sigma}_{\mu} \chi_1)+{\delta \over 2 m}(\overline{\chi}_2
\overline{\sigma}_{\mu} \chi_2-\overline{\chi}_1
\overline{\sigma}_{\mu} \chi_1).
\end{equation}
Because $\delta /m \sim 10^{-6}$, we ignore the second term, and find
that the only way for $\chi_1$ to scatter coherently off of nuclei is to
make a transition into the heavier $\chi_2$ state.  This inelastic process
will dominate relative to the elastic, spin dependent scattering
provided that the coherence enhancement, which
gives a factor $\sim
A^2 \sim 5 \cdot 10^3$ in the cross section, overcomes
than the suppression due to the inelasticity.  In this case,
the rate can depend sensitively on the mass of the target nucleus, as desired.
%

\subsection{Sneutrino Dark Matter}
\label{subsection:model}
Interestingly enough, a suitable candidate for inelastic dark matter
has already been discussed in the literature.
In supersymmetric theories with lepton number violation, the LSP can be
a real component of the sneutrino \cite{Hall:1998ah,
  Arkani-Hamed:2000bq, Arkani-Hamed:2000kj}:
a Lagrangian term $-\Delta ^2 \tilde{\nu} \tilde{\nu}+h.c.$ 
lifts the degeneracy
between the sneutrino's odd and even CP eigenstates $\tilde{\nu}_-$ and 
$\tilde{\nu}_+$.  This splitting prevents elastic
scattering of the lightest state, $\tilde{\nu}_-$, 
off of nuclei through $Z$ exchange\footnote{There are contributions 
that will induce an elastic scattering, for instance from Higgs 
exchange, but these are all small and can be ignored for our purposes 
here.}. There is still the
challenge of achieving a cosmologically 
interesting relic abundance - for an
ordinary $~100$ GeV sneutrino $\Omega_{\tilde{\nu}}$ comes out too small.  
In \cite{Hall:1998ah}, this problem was resolved by taking the mass splitting
between $\tilde{\nu}_-$ and $\tilde{\nu}_+$ to be large enough
to prevent coannihilation via an s-channel $Z$ in the early universe, 
$\delta> 5$ GeV, leading to a
radiatively generated neutrino neutrino mass $m_{\nu}>$ 5 MeV.   
Different approaches were taken in the models of
\cite{Arkani-Hamed:2000bq, Arkani-Hamed:2000kj, Borzumati:2000mc}. 
These models feature standard model singlet scalars $\tilde{n}$ that
are kept light by a global 
symmetry \cite{Arkani-Hamed:2000bq, Arkani-Hamed:2000kj} in analogue 
to the Giudice-Masiero solution to the $\mu$ 
problem \cite{gm}, 
or by a gauged B-L symmetry \cite{Borzumati:2000mc}.
The singlet states mix with ordinary sneutrinos through weak scale $A$
terms, so that the gauge interactions of the mass eigenstates are suppressed
by mixing angles.  This suppression allows for an
interesting relic abundance even for values of $\delta$ too small to
prevent coannihilation between $\tilde{\nu}_-$ and $\tilde{\nu}_+$.

For concreteness we will specialize to the model of 
\cite{Arkani-Hamed:2000kj}.  
The global symmetry that prevents a tree level mass for the singlet
$\tilde{n}$ states is broken by the vev of a spurion $X$ that also breaks supersymmetry. 
We assume that the $A$ and $F$ components of $X$ both have
intermediate scale vevs: $\langle A_X \rangle \sim \sqrt{\langle F_X
  \rangle} \sim m_I \sim \sqrt{v M_{Pl}}$.  The spurion couples to
the neutrino and singlet superfields according to
\begin{equation}
\mathcal{L}\supset {1 \over M_{Pl}}\left[XLNH_u \right]_F+{1\over M_{Pl}}[X^\dagger
NN\left(1+{X^\dagger X \over M^2_{Pl}}+\ldots \right)]_D+h.c.
\label{eq:snucouplings}
\end{equation}
The operators of (\ref{eq:snucouplings}) can be justified by ordinary R
parity (under which $N$ is odd and $X$ is even), \
together with an R symmetry where $N$
has R charge 2/3, $X$ has charge 4/3, and $L$ and $H_u$ have R charge
0. As discussed in \cite{Arkani-Hamed:2000kj}, at tree level
(\ref{eq:snucouplings}) yields a neutrino mass matrix whose light
eigenvalue is $\sim v^2/M_{Pl}$.  However, (\ref{eq:snucouplings})
also contains 
\begin{equation}
\mathcal{L}\supset -A \tilde{l} \tilde{n} h_u - 
\Delta^2(\tilde n\tilde n+h.c.),
\label{eq:snucouplings2}
\end{equation} 
with $A$ roughly weak scale and $\Delta^2 \sim m_I^5/M_{Pl}^3$.  
These interactions radiatively
induce a Majorana mass for the left-handed neutrino
\begin{equation}
m_\nu\sim{g^2\over 384\pi^2}{v^{3/2}\over M_{Pl}^{1/2}}
\end{equation} 
that is larger than that obtained from the tree-level seesaw, and moreover,
in roughly the correct range for explaining the
atmospheric neutrino anomaly.

For our present purpose, however, the impact of (\ref{eq:snucouplings2}) 
on the scalar masses is what matters most.  Neglecting the small
lepton number violating mass parameter $\Delta$, the sneutrino
mass-squared matrix is
\begin{equation}
\mathcal{L}\supset-\left(\begin{array}{cc}\tilde{\nu}^* & n
  \end{array}\right)\left
(\begin{array}{cc}m_L^2 
&{1 \over \sqrt{2}} A v\sin\beta  
\\{1 \over \sqrt{2}} A v\sin\beta 
& m_R^2
  \end{array} \right)
\left
(\begin{array}{c}
\tilde{\nu} \\ n^*
  \end{array} \right).
\label{eq:matrix}
\end{equation}
The $A$ term coupling induces a mixing between $\tilde{\nu}$ and
$\tilde{n}$, yielding a lighter mass eigenstate
\begin{equation}
\tilde{\nu}_1=-\tilde{\nu} \sin\theta+\tilde{n}^* \cos\theta.
\end{equation}
The coupling of $\tilde{\nu}_1$ to the $Z$ boson is thus suppressed by $\sin^2 \theta$.
The lepton number violating parameter $\Delta^2$ lifts the degeneracy
between the CP-even and CP-odd components 
of $\tilde{\nu}_1$, leading to a small 
mass difference
\begin{equation}
\delta \simeq 2 \cos^2 \theta {\Delta^2 \over m_1}.
\end{equation}
For this splitting to resolve the conflict between CDMS and DAMA,
one needs $\delta \sim 50-100$ keV, roughly\footnote{The lifetime
  of $\tilde{\nu}_+$ is $\tau \simeq \left({1/3\over \sin\theta}\right)^4 \left({{\rm 100\;\;
      keV}\over \delta}\right)^5 (4\cdot 10^2 {\rm \;\;years})$, so
for the mass
  splittings and mixing angles of interest, it is safe to assume that
  only $\tilde{\nu}_-$ is present today. Photons can be produced in
  these decays, but the decays take place before recombination for the
  parameters of interest, and the photons are soft enough to render
  negligible the effect on the CMBR spectrum.}.  For a 100 GeV
sneutrino, this implies $[X^\dagger X X^\dagger]_D \sim m_I^5 
\sim (3 \cdot 10^{10} {\rm\;\; GeV})^5$, corresponding to a reasonable value for the
intermediate scale.  

\begin{figure}
\centerline{\psfig{file=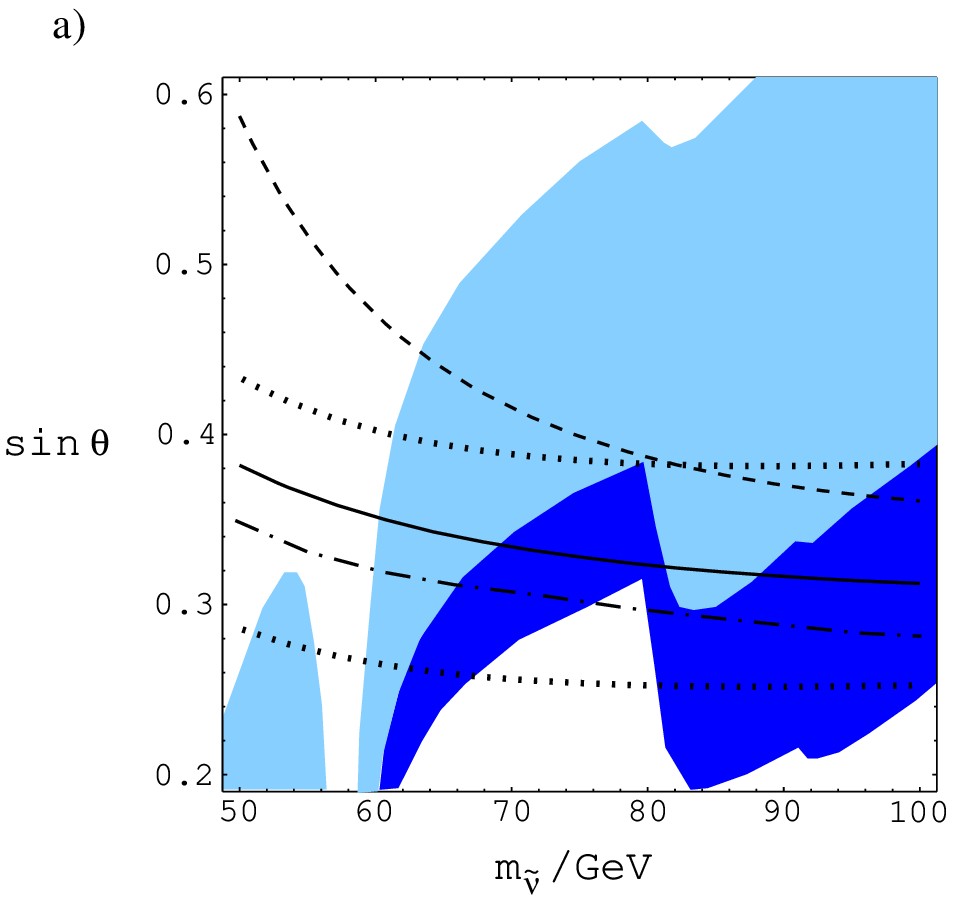,width=0.4\textwidth,angle=0}\psfig{file=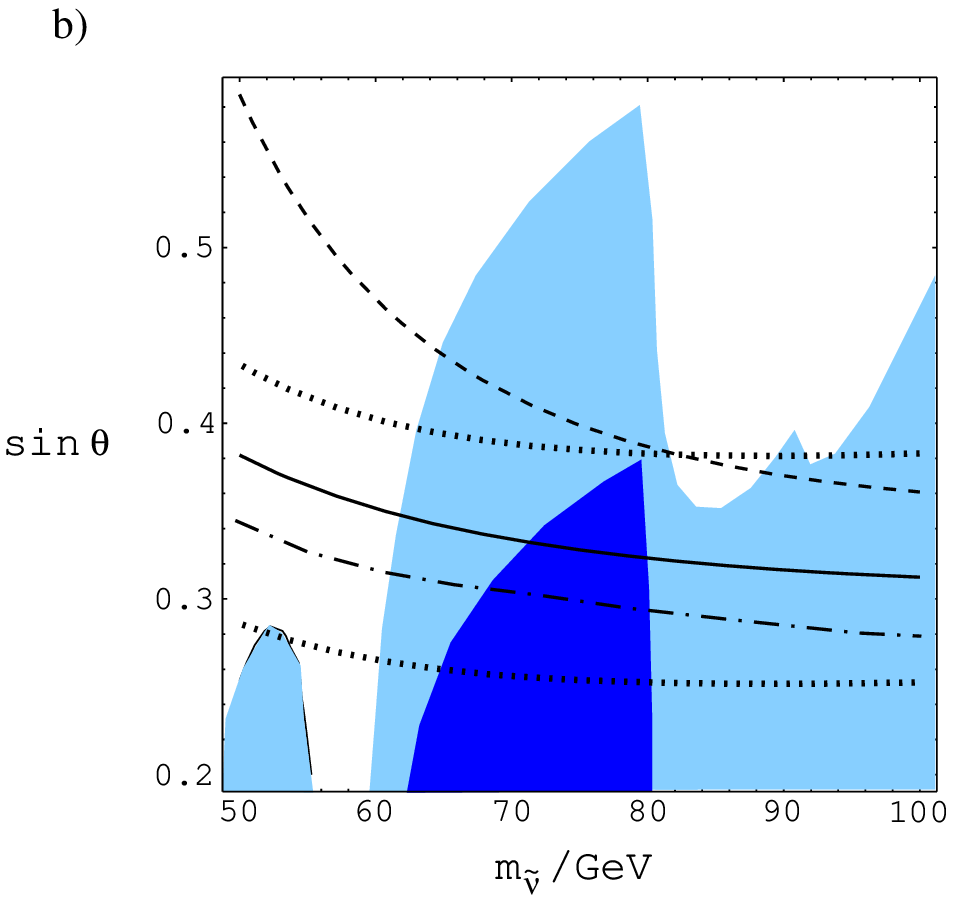,width=0.4\textwidth,angle=0}}
\centerline{\psfig{file=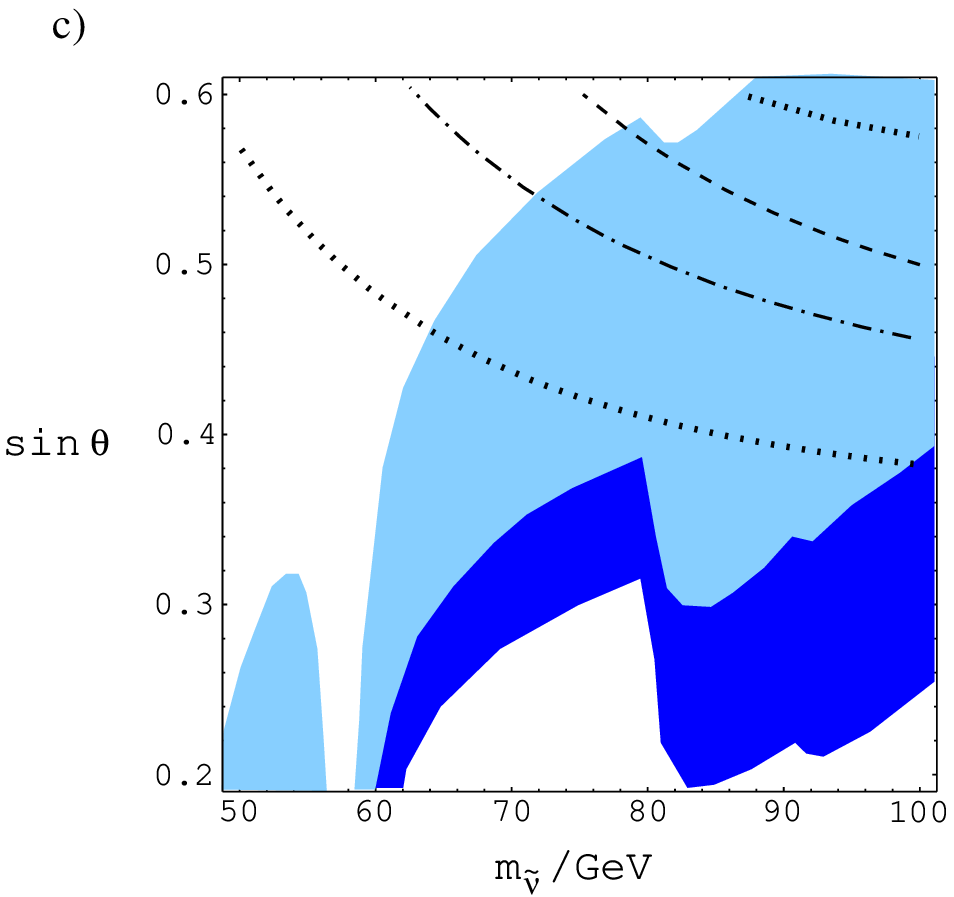,width=0.4\textwidth,angle=0}\psfig{file=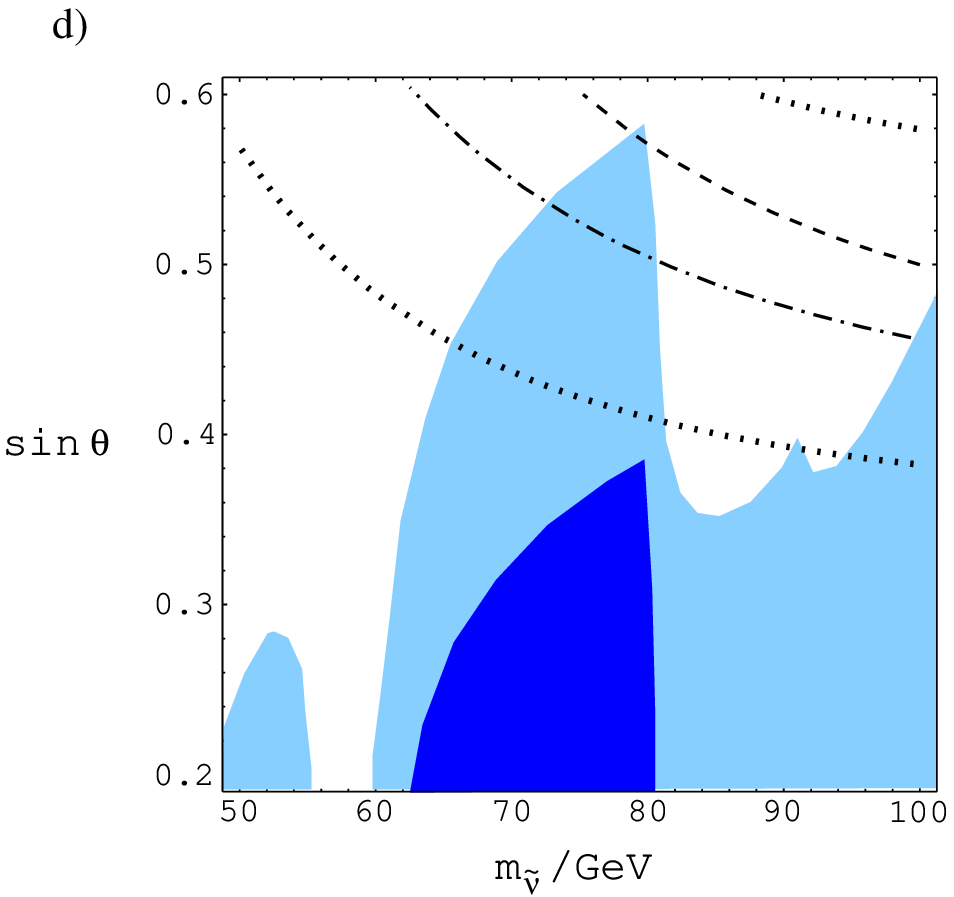,width=0.4
\textwidth,angle=0}}
\caption{For the sneutrino dark matter case, 
regions that satisfy the direct detection requirements of
  section 3, plotted along with filled contours of
  $\Omega_{\tilde{\nu}} h^2$.  The lighter shaded region corresponds
  to $.05<\Omega_{\tilde{\nu}} h^2<.3$ and the darker shaded region corresponds
  to $.3<\Omega_{\tilde{\nu}} h^2<.64$.  The region between the dotted
contours has an integrated signal in the 2 - 6 keV range consistent
with the DAMA $3\sigma$ region.  The solid line gives the CDMS
constraint, the dashed line gives the limit from the absence of
signal in the high energy bins at DAMA, and 
the dot-dashed line gives the constraint 
arising from Xe pulse shape analysis data (regions below these
lines are allowed).   We take $\delta=50$ keV for 
a) and b) and $\delta=100$ keV for c) and d).  For a) and c), we use $A=25$
  GeV while for  b) and d), we take $A=50$ GeV. 
For each plot we take $\tan\beta=50$, $m_h=115$ GeV, 
  and a bino mass of $300$ GeV, with the assumption of GUT
  unification of gaugino masses.}
\label{fig:snuplots}    
\end{figure}

To explore the feasibility of this scenario, we  
apply the same criteria used in section 3 to establish consistency
with CDMS and DAMA for  
$\delta=50$ and 100 keV, and display the allowed regions in the
($m_{\tilde{\nu}}$, $ \sin\theta$) plane.  
Note that because the scattering off of nuclei is suppressed both by
the inelasticity of the reaction and by a $\sin^4\theta$ factor, the
ability to obtain a large enough signal at DAMA depends crucially on
the fact that ordinary sneutrinos give a signal roughly three orders
of magnitude above present bounds. We also calculate
the relic abundance as a function of $m_{\tilde{\nu}}$ and
$\sin \theta$ using standard methods.  The results shown in figures
\ref{fig:snuplots}a)-d) indicate that there are indeed 
regions of parameter space featuring interesting relic abundances and
acceptable direct detection rates.  In the early universe, the
efficiency of annihilation processes that occur via s-channel Higgs
exchange, such as $\tilde{\nu}_- \tilde{\nu}_- \rightarrow
b\overline{b},\; ZZ,\; W^+W^-$, is sensitive to the size of the trilinear
scalar coupling $A$, leading to the dependence of the relic abundance
on $A$ evident in the figures.

Just as one specific illustrative example, consider the parameters
$m_{\tilde{\nu}}=70$ GeV, $\delta=70$ keV, and $\sin^4\theta=1/70$. 
This choice of $\sin\theta$ leads to 
an interesting relic abundance for a broad range of SUSY parameters.
For this choice of $m_{\tilde{\nu}}$, $\delta$, and
$\sin\theta$, we calculate a mean of less than 2 events at CDMS and 
satisfy the constraints from the Xe pulse shape analysis.
Moreover, 
as shown in table \ref{table:snutable}, 
\begin{table}
\begin{center}
\begin{tabular}{|c|c|c|}\hline
${\rm Energy/keV}$&\multicolumn{2}{c|}{$S_{m,k}$(cpd/kg/keV)}\\
\cline{2-3}
 & DAMA best& inelastic $\tilde{\nu}$ \\ \hline
2-3&.027 & .027 \\
3-4& .013& .013\\
4-5& .005& .006\\
5-6& .002& .002 \\
\hline
\end{tabular}
\end{center}
\caption{$S_{m,k}$ values obtained using DAMA's best fit point,
$\sigma=7.2\cdot 10^{-6}$ pb and $m_\chi= 52$ GeV for the standard
WIMP case, 
and values obtained taking 
$\delta =70$ keV, $m_{\tilde{\nu}}=$ 70 GeV, and $\sin^4\theta=1/70$
for the sneutrino inelastic dark matter case.}
\label{table:snutable}
\end{table}
the values of $S_{m,k}$ we
obtain for DAMA in the 2-6 keV energy range are nearly identical to
those we obtain in the elastic case using DAMA's best fit point
$\sigma_n=7.2\cdot 10^{-6}$ pb and $m_\chi= 52$ GeV.   

\subsection{Indirect detection}
As dark matter passes through the sun, it can scatter off of nuclei 
and be captured in the sun's potential well \cite{indirect}.
After a significant 
amount of dark matter has been captured, 
it can annihilate into other particles. 
If muon neutrinos are produced, those that reach the Earth can 
produce high energy muons through charged-current
interactions.  A number of experiments have attempted to direct WIMP
matter indirectly by looking for these upward-going muons, leading to
a current limit on their flux of $10^{-14}$ cm$^{-2}$ s$^{-1}$
\cite{Okada:2000ve,Michael:1998ky,LoSecco:1987fu}.

Even within the model of section \ref{subsection:model}, the expected
flux of upward-going muons is quite uncertain, for a number of
reasons.  First, the capture rate in the sun is sensitive to the parameters
$m_{\tilde \nu}$, $\sin\theta$ and $\delta$.  Second, if the captured
sneutrinos annihilate directly into neutrinos, 
the flavor of the lightest sneutrino determines what flavor 
of neutrino is produced, and details of the neutrino masses and 
mixings impact the flavor of the neutrino detected at the 
Earth. Third, cosmological uncertainties mentioned in section
\ref{subsection:cosmun} 
can change the preferred region 
of $\sigma_{n}$ and thus the capture rate. Finally, relatively minor 
extensions to the model of section \ref{subsection:model} can also complicate matters.
As a consequence of these various sources of uncertainty, 
indirect techniques do not  rule out inelastic sneutrino dark matter. 
However, they do impose strong constraints,
as broad regions of parameter space lead to signals 
above experimental bounds. Moreover, 
indirect detection experiments offer the 
strong possibility of detection if the bound on the muon flux improves 
considerably \cite{Halzen:1999jy}.

One might expect that the same
inelasticity that suppresses the signal at CDMS should be even more
effective in suppressing the capture rate by the sun, which is mainly
composed of relatively light nuclei.  In fact, this is typically not
the case.  Because particles passing through the sun are unusually
energetic (the escape velocity at the surface of the sun is
much larger than the average velocity of a halo particle),
the inelasticity is {\em less} relevant in the sun than at direct
detection experiments.

In what follows, we have followed \cite{gould} in calculating solar
capture rates and the 
induced muon flux, but have modified the approach to approximate 
the suppression of the capture rate due to the inelasticity (see
Appendix).  This suppression depends on the mass of the nucleus.  For
example, for $\delta=100$ keV and $m_{\tilde{\nu}}=100$ GeV, we find a factor
$\sim 20$ suppression for scattering off of oxygen in the sun, and a
factor $\sim 2$ suppression for scattering off of iron.

We will separately consider first, the case in which
the sneutrinos cannot annihilate into W's, and second, the case
in which they can.
For $m_{\tilde{\nu}}<m_W$, sneutrinos in the sun typically 
annihilate dominantly to neutrinos
via t-channel neutralino exchange.
If we neglect cosmological uncertainties, 
we find that for values of $m_{\tilde{\nu}}$, $\delta$ and $\sin\theta$
that lead to interesting relic abundances and consistency with CDMS and
DAMA, the flux of neutrinos produced is quite large. If these are all 
muon flavor, we would expect a flux of upward-going muons of at least 
$\sim 6 \times 10^{-13}{\rm cm^{-1}s^{-1}}$, in conflict with experimental 
results. 
On the other hand, if the sneutrinos annihilate into electron
neutrinos that do not oscillate into muon neutrinos, bounds from
direct detection are evaded entirely.  If this is the case, only 
direct detection experiments will be able to yield a positive signal.

Finally, we note that there
are specific parameter choices for which the dominant
annihilation of sneutrinos in the sun is through s-channel Higgs to
$b \overline{b}$.  We find that this allows the flux of upward-going
muons to be as small as a factor of $\sim 2$ 
above current limits for parameters
that yield an acceptable abundance and acceptable direct detection signals. 

With cosmological uncertainties included, more scenarios are allowed. 
Relatively small variations in $v_{rms}$ can accomodate factors of 
two, such as if the dominant annihilation is into $b \overline b$.
Direct annihilation into muon neutrinos would require a more specious 
conspiracy of errors. For instance, if the solar system were 
presently in an anomalously high density region of the galaxy arising 
from substructure, and if $v_{rms}$ were $3\sigma$ above the value we 
have used, experiments could accomodate as much as one-third of the 
neutrinos produced being muon flavored. This seems quite unlikely, but is, at 
least in principle, still allowed.

However, for these lighter sneutrinos, if indirect detection experiments 
improve by an order of magnitude, they will be 
able to probe almost all of the parameter space, even accounting for a 
broad class of cosmological uncertainties, and situations where 
there is annihilation to $b \overline b$.

For heavier sneutrinos ($m_{\tilde{\nu}}>m_W$), the dominant
annihilation processes in the sun can easily be s-channel Higgs exchange
to W's and Z's.  In  this case we find that it is possible to
reduce the expected signal at direct-detection experiments to a factor
of $\sim 3$ above current limits for parameters consistent with DAMA,
CDMS, and $\Omega_{\tilde{\nu}} h^2 \sim .1$.  
These heavier sneutrinos are less affected by cosmological 
uncertainties, but these uncertainties still make it impossible to 
rule out this scenario. Future
improvements in indirect detection could rule out
this region of parameter space, especially as the experimental
signal is less sensitive to the flavor of 
the lightest sneutrino than for the case of $m_{\tilde \nu}<m_{W}$.

Of course, this discussion applies only to the model of section \ref{subsection:model}, 
and one can consider modifications to the model 
that suppress the indirect detection signal.
The premise of the model is that light standard model singlets are natural. 
Given this, 
if we add to that model another standard model singlet $\eta$,
with the same R-charge as the right handed neutrino $N$, but opposite 
R-parity, we expect a superpotential interaction $\eta NN$. Then 
through t-channel $\eta$ exchange, $\tilde{\nu}$'s 
can annihilate to right-handed 
neutrinos. If these decay dominantly into muons or electrons (rather
than tau's) and off-shell W's, we find
that it is possible to bring the flux of upward going muons 
induced by the W decay products down to current 
limits\footnote{In such a scenario, the relic abundance is modified, 
but it is still possible to have $\Omega_{\tilde \nu} h^{2} \sim 
0.1$.}.  Future indirect experiments would still likely be able to 
see the decay products of these right handed neutrinos. 

This is just one example of a modification to the model which 
diminishes the signal, and there may be others, but such uncertainty is 
difficult to quantify.
While indirect experiments offer a good opportunity to test 
specific models and regions of parameter space, there is an excellent 
likelihood that upcoming direct detection experiments will be able to 
determine whether inelastic dark matter is the resolution of the 
conflict between DAMA and CDMS.

\section{Future Experiments}
In the inelastic dark matter scenario, the boundaries of the DAMA
 preferred region are not far from the current limits 
from CDMS. Planned experiments should be able to cover the existing 
DAMA region. Most important are planned improvements to 
germanium experiments, and the CRESST experiment, which will use the
heavy element tungsten.

CDMS will soon be moving to the Soudan mine, and should 
be able to improve its limits by at least two orders of magnitude 
\cite{Sadoulet:2000rq}. The GENIUS Ge experiment \cite{Klapdor-Kleingrothaus:1999hk} 
should go well below that, likely allowing both to test much of the 
preferred regions discussed in section \ref{sec:sigs}. 

There is a caveat in this statement: in generating the plots of figure
\ref{fig:regions}, 
we neglected to include the effect of a finite galactic 
escape velocity. This was a harmless simplification 
for our purposes there, 
because the effects at DAMA due to the finite galactic escape velocity are 
relatively minor. The effects can be much larger at CDMS.

Recall that the requirement for scattering is
\begin{equation}
        \delta < {\beta^{2} \over 2} \frac{m_{N} m_{\chi}}{m_{N}+m_{\chi}}.
        \label{eq:redundant}
\end{equation}
This constraint is particularly stringent for light candidates. For instance, 
with $m_{\chi}=50\gev$ and $v_{esc}=650{\rm km/s}$ equation 
(\ref{eq:redundant}) tells us that only for $\delta < 122 \kev$
can one hope to obtain any signal at all at a germanium detector 
(recall that the 
highest velocity of particles incident on the earth is $v_{esc}+v_{\odot}$). 
Thus, the higher $\delta$ regions may not be testable at CDMS.

For heavier candidates, the finite galactic escape velocity
is not especially important, even at CDMS. 
With a galactic escape velocity of $650{\rm km/s}$, and 
$m_{\chi}=100\gev$, the cutoff for $\delta$ is $172\kev$. 
On the other hand, the galactic escape 
velocity is not a particularly well known quantity, and if instead
we take $v_{esc}=450{\rm 
km/s}$, the cutoff for $\delta$ is only $102 \kev$.

These uncertainties make the  
CRESST experiment \cite{DiStefano:2000gu}, using tungsten, especially 
significant. Because tungsten ($A=183$) is heavier than iodine 
($A=127$), given adequate exposure time, CRESST should cover the DAMA 
preferred region, irrespective of cosmological uncertainties.

A very real possibility is that both germanium and tungsten
experiments {\em will} have 
signals, which, when interepreted as elastic scatterings, would be 
inconsistent with one another. The most striking possibility of all is 
a spectrum deformation at the germanium detectors, as 
discussed in section \ref{sec:sigs}. 
If CDMS were to see an excess of events in the $30-70 \kev$ region, 
but no excess below $30 \kev$, it would be a compelling signature of this 
scenario.

\section{Conclusions}
If in fact the majority of the matter of the universe is non-baryonic, 
the attempt to determine its nature is one of the most exciting 
endeavors of modern cosmology. Existing dark matter searches have 
already begun to probe interesting regions of parameter space for 
candidate particles such as neutralinos and axions. 

The positive result from the DAMA experiment is difficult to a
understand in terms of these candidates, as it is in seeming conflict 
with constraints arising from the CDMS experiment. 
We have seen that this conflict vanishes if we allow for the 
possibility 
that the dark matter particle can only scatter inelastically. 

We have shown that the sneutrino, when mixed with a 
singlet scalar with weak lepton number violation, is a viable
candidate for inelastic dark matter. 
The regions of parameter space 
which give an interesting relic 
sneutrino abundance 
overlap with the regions which give a positive DAMA signal.  Indirect
detection experiments tightly constrain models of sneutrino dark
matter, but do not rule them out.

Even absent a particular model, 
we find it interesting that such a simple modification of the 
dark matter's properties can give remarkably different predictions, 
including the suppression of a signal at CDMS. We consider these 
results sufficiently interesting as to warrant an analysis of 
the full DAMA data set should the 
raw data become available. 

\vskip 0.3in

{\noindent \Large \bf Acknowledgments}
\vskip 0.15in
We thank L. Hall, W. Haxton, C. Hogan, and T. Quinn for useful
discussions, and H. Murayama and A. Nelson
for reviewing the paper and providing valuable comments. 
 
\appendix
\section{Appendix}
Here we describe how we approximate the suppression 
of the rate of relic capture by the sun due to the inelasticity of the
scattering. Ignoring nuclear form factors, the scattering probability
for a given relative velocity $w$ is equally distributed between the
minimum and maximum
nuclear recoil energies $\Delta E_{min}$ and $\Delta E_{max}$ (these
parameters depend on the nucleus mass, the relic mass $m$, $w$, and $\delta$).
Ordinarily, the low-energy scattering cross section is independent of
$w$, but in the inelastic case there is an additional phase space
factor $\sqrt{1-2\delta/(\mu w^2)}$, where $\mu$ is the reduced mass.
Capture only occurs when  $\Delta E>\Delta E_{capture}\equiv 1/2\left( mw^2-(m+\delta)
v_{esc}^2(r)\right)-\delta$ holds.  Here $v_{esc}(r)$ is the (position-dependent)
escape velocity, which we approximate as \cite{gould}
\begin{equation}
v_{esc}(r)=v_c^2-{M(r)\over M_{\odot}}(v_c^2-v_s^2),
\end{equation} 
where $v_c=1354$ km/s,  $v_s=795$ km/s, and $M(r)$ is the mass
contained within the radius $r$.  The capture rate off of a
given species of nuclei is then proportional to
\begin{equation}
\int_0^{R_{\odot}}\!dr \; r^2 \rho(r) \int_{v_{esc}}^{\infty} \! dw \;
w^3 e^{-{(w^2-v_{esc}^2)\over v_0^2}}\sqrt{1-2\delta/(\mu
  w^2)}\left( {\Delta E_{max}-\Delta E_{capture}\over
  \Delta E_{max}-\Delta E_{min}}\right),
\end{equation}
where $\rho (r)$ is the mass density of the species and $v_0$ is the
rotational speed of the local standard of rest.
We calculate this factor (which does not account for form factor suppressions)
in the elastic ($\delta=0$) and inelastic cases to estimate the
suppression coming from the inelasticity.  We then obtain capture
rates  by multiplying this suppression with the rate obtained for the
elastic case using the formulae of \cite{gould} (which do include form
factor suppressions).


\begin{thebibliography}{99}
        
\bibitem{Trimble:1987ds}
V.~Trimble,
Ann.\ Rev.\ Astron.\ Astrophys.\  {\bf 25}, 425 (1987).


\bibitem{Alcock:2000ph}
C.~Alcock {\it et al.}  [MACHO Collaboration],
Astrophys.\ J.\  {\bf 542}, 281 (2000)
[astro-ph/0001272].


\bibitem{Lee:1977ua}
B.~W.~Lee and S.~Weinberg,
Phys.\ Rev.\ Lett.\  {\bf 39}, 165 (1977).

\bibitem{Baudis:2001ph}
L.~Baudis, A.~Dietz, B.~Majorovits, F.~Schwamm, H.~Strecker and H.~V.~Klapdor-Kleingrothaus,
Phys.\ Rev.\  {\bf D63}, 022001 (2001)
[astro-ph/0008339].

\bibitem{Abusaidi:2000wg}
R.~Abusaidi {\it et al.}  [CDMS Collaboration],

Nucl.\ Instrum.\ Meth.\  {\bf A444}, 345 (2000)
[astro-ph/0002471].

\bibitem{Bernabei:2000qi}
R.~Bernabei {\it et al.}  [DAMA Collaboration],
Phys.\ Lett.\  {\bf B480}, 23 (2000).

\bibitem{Ullio:2000bv}
P.~Ullio, M.~Kamionkowski and P.~Vogel,
hep-ph/0010036.

\bibitem{Hall:1998ah}
L.~J.~Hall, T.~Moroi and H.~Murayama,
Phys.\ Lett.\  {\bf B424}, 305 (1998)
[hep-ph/9712515].

\bibitem{Freese:1988wu}
K.~Freese, J.~A.~Frieman and A.~Gould,
Phys.\ Rev.\  {\bf D37}, 3388 (1988).

\bibitem{form} R.H.~Helm, Phys. Rev. {\bf D104},1466 (1956); J.~Engel,
  Phys. Lett. {\bf B264}, 114 (1991).

\bibitem{Moore:1999wf}
B.~Moore, S.~Ghigna, F.~Governato, G.~Lake, T.~Quinn, J.~Stadel and P.~Tozzi,
astro-ph/9907411.

\bibitem{Vergados:2000rh}
J.~D.~Vergados,
Phys.\ Rev.\ D {\bf 62}, 023519 (2000)
[astro-ph/0001190].

\bibitem{Bernabei:1996vj}
R.~Bernabei {\it et al.},
Phys.\ Lett.\  {\bf B389}, 757 (1996).

\bibitem{Bernabei:1998ad}
R.~Bernabei {\it et al.},
Phys.\ Lett.\ {\bf B436}, 379 (1998).


\bibitem{Arkani-Hamed:2000bq}
N.~Arkani-Hamed, L.~Hall, H.~Murayama, D.~Smith and N.~Weiner,
hep-ph/0006312.

\bibitem{Arkani-Hamed:2000kj}
N.~Arkani-Hamed, L.~Hall, H.~Murayama, D.~Smith and N.~Weiner,
hep-ph/0007001.

\bibitem{gm} G.F.~Giudice and A.~Masiero, Phys. Lett. {\bf B206} (1988) 480.

\bibitem{Borzumati:2000mc}
F.~Borzumati and Y.~Nomura,
hep-ph/0007018.

\bibitem{indirect}J.~Silk, K.~Olive and M.~Srednicki,
  Phys. Rev. Lett. 55 (1985) 257; L.M.~Krauss, K.~Freese, D.N.~Spergel
  and W.H.~Press, Astrophys J. 299 (1985) 1001.


\bibitem{Okada:2000ve}
A.~Okada  [Super-Kamiokande Collaboration],
astro-ph/0007003.

\bibitem{Michael:1998ky}
D.~G.~Michael  [MACRO Collaboration],
{\it Prepared for 29th International Conference on High-Energy Physics (ICHEP 98), Vancouver, British Columbia, Canada, 23-29 Jul 1998}.

\bibitem{LoSecco:1987fu}
J.~M.~LoSecco {\it et al.},
Phys.\ Lett.\  {\bf B188}, 388 (1987).

\bibitem{Halzen:1999jy}
F.~Halzen {\it et al.}  [AMANDA Collaboration],
{\it Prepared for 26th International Cosmic Ray Conference (ICRC 99), Salt Lake City, UT, 17-25 Aug 1999}.

\bibitem{gould} A.~Gould, Astrophys. J. {\bf 388}, 338 (1992);
G.~Jungman, M.~Kamionkowski, and K.~Griest, Phys. Rep.{\bf 267}, 195 (1996).

\bibitem{Sadoulet:2000rq}
B.~Sadoulet,
in {\it Proc. of the 19th Intl. Symp. on Photon and Lepton Interactions at High Energy LP99 } ed. J.A. Jaros and M.E. Peskin,
eConf {\bf C990809}, 687 (2000).

\bibitem{Klapdor-Kleingrothaus:1999hk}
H.~V.~Klapdor-Kleingrothaus {\it et al.}  [GENIUS Collaboration],
hep-ph/9910205.

\bibitem{DiStefano:2000gu}
P.~Di Stefano {\it et al.},
hep-ex/0011064.



        
\end{thebibliography}
\end{document}